\documentclass[aps,prl,twocolumn,superscriptaddress]{revtex4-1}

\usepackage[compat=1.0.0]{tikz-feynman}
\usepackage{amssymb}
\usepackage{multirow}
\usepackage{bm}
\usepackage{changes}
\usepackage{amsmath}
\usepackage{graphicx}
\usepackage{epstopdf}
\usepackage{subfigure}
\usepackage{natbib}
\usepackage{epsfig}
\usepackage{amsfonts}
\usepackage{mathrsfs}
\usepackage{braket}
\usepackage[toc,page,title,titletoc,header]{appendix}
\usepackage[colorlinks,linkcolor=blue,citecolor=blue,anchorcolor=blue]{hyperref}
\usepackage{dsfont,amsthm,amsbsy}

\usepackage{fancyhdr}

\usetikzlibrary{quotes,angles}
\newcommand{\obtuseangle}{\kern.08em
\begin{tikzpicture}
    \draw coordinate (a) at (0.14,0);
    \draw coordinate (b) at (0,0);
    \draw coordinate (c) at (-.12,0.18);
    \draw (a) -- (b) -- (c) pic [draw=black]{} ;
\end{tikzpicture}%
\kern.08em%
}

\begin{document}
\title{High Temperature Superconductivity in a Lightly  Doped Quantum Spin Liquid}
\author{Hong-Chen Jiang}
\email{hcjiang@stanford.edu}
\affiliation{Stanford Institute for Materials and Energy Sciences, SLAC National Accelerator Laboratory and Stanford University, Menlo Park, California 94025, USA}

\author{Steven A. Kivelson}
 \affiliation{Department of Physics, Stanford University, Stanford, California 94305, USA}

\date{\today}
\begin{abstract}
We have performed density-matrix renormalization group studies of a square lattice $t$-$J$ model with small hole doping, $\delta\ll 1$, on long 4 and 6 leg cylinders.  We include frustration in the form of a second-neighbor exchange coupling, $J_2 = J_1/2$, such that the undoped ($\delta=0$) ``parent'' state is a quantum spin liquid. In contrast to the relatively short range superconducting (SC) correlations that have been observed in recent studies of the 6-leg cylinder in the absence of frustration, we find  power law SC correlations with a Luttinger exponent, $K_{sc} \approx 1$, consistent with a strongly diverging SC susceptibility, $\chi \sim T^{-(2-K_{sc})}$ as the temperature $T\to 0$.  The spin-spin correlations - as in the undoped state - fall exponentially suggesting that the SC ``pairing'' correlations evolve smoothly from the insulating parent state.
\end{abstract}

\maketitle

Although the physics of the cuprate high temperature superconductors is surely complex, there are a variety of reasons\cite{anderson87,emery87,scalhubbard} to believe that the ``essential''\cite{complexity} physics is captured by the two-dimensional (2D) Hubbard model or its close relatives. 
To begin with, as is the case in the cuprates, in an appropriate regime of parameters, 
 the Hubbard model on a square lattice with $n=1$ electrons per site  exhibits an  undoped ``parent'' state 
that is a Mott insulating antiferromagnet. 
However, two key theoretical issues concerning this proposition  remain unsettled: 1) Does d-wave superconductivity (SC) ``robustly'' arise in this model upon light doping, i.e. for $0 < \delta\equiv (1-n) \ll 1$.  2)  If so, how does it arise (i.e. what is the ``mechanism'') and under what circumstances (e.g. does it depend on specific features of the band structure)?

For parametrically small values of the Hubbard $U\ll W$ (where $W$ is the bandwidth), it is possible to establish\cite{Raghu2010} that such a superconducting state arises, but here (except under extremely fine tuned circumstances in which the Fermi surface is perfectly nested) the undoped state at $n=1$ is also superconducting, and the superconducting $T_c$ is exponentially small in units of $W$.  For intermediate $U \sim W$, no controlled analytic approach exists, but calculations based on a variety of physically motivated approximations  \cite{gull, maier,andrey} yield  results suggestive of values of $T_c$ as large as $T_c \sim W $ (where the proportionality is a number of order 1 but may be small, e.g. $\sim (2\pi)^{-2}$). This was further supported by density-matrix normalization group (DMRG) studies of the Hubbard and $t$-$J$ models on 4-leg square cylinders.\cite{Jiang2018tJ,Jiang2019Hub,Jiang2020Hub,Jiang2020tJ,Chung2020} However, recent\cite{Simons2015,whitenew} DMRG calcualtions on 6-leg square cylinders, as well as variational Monte Carlo\cite{sorella} calculations on 2D models, have called this proposition into question. Specifically, the tendency of a doped antiferromagnet to phase separation\cite{emeryandlin,sorella} or to charge-density wave (CDW) formation\cite{zaanenstripes,shultzstripes1,machidastripes1,whitescalapino4leg,Simons2015,whitenew,Qin2020,Jiang2020Hub} appear to play a much more dominant role in the physics at small $\delta$ than accounted for by most approximate approaches.

One attractive notion that was suggested early on is that high temperature superconductivity could arise naturally\cite{anderson87,KRS,laughlinsc,kotliar,balentsandnayak,vanilla,Song2021} under circumstances in which the  insulating parent state is a quantum spin liquid (QSL) rather than an ordered antiferromagnet.  In particular, a QSL with a gap (even a partial gap with nodes), can in some sense be thought of as a state with pre-existing Cooper pairs  but with vanishing superfluid stiffness.  Then, upon light doping, one might naturally expect SC with a gap scale that is inherited from the QSL (i.e. evolves continuously as $\delta\to 0$) and with a superfluid stiffness - that rises with $\delta$.

In the present paper, we explore the possibility of SC in a doped spin liquid using DMRG to treat the $t$-$J$ model (a proxy for the Hubbard model) on cylinders of circumference 4 and 6.  A number of studies of the spin-1/2 Heisenberg model on the square lattice with first and second neighbor exchange couplings, $J_1$ and $J_2$, have led to a consensus\cite{figandsondhi,Capriotti2001,Jiang2012,Hu2013,Gong2014,Morita2015,Wang2016,Wang2018} that there is a QSL phase in the range of $0.46 < J_2/J_1 < 0.52$.\cite{Wang2018}  In this range, DMRG on cylinders of circumference up to $L_y=10$ show a pronounced spin-gap and exponentially falling spin-spin correlations with a  correlation length $\xi_s$ considerably smaller than $L_y$.\cite{Jiang2012,Gong2014} However, there is still some debate about whether this gap persists in the 2D limit, or if instead the QSL phase has a gapless nodal spinon spectrum.

Here, we study the model with $J_2/J_1=0.5$, and correspondingly we take the ratio of nearest to next-nearest neighbor hopping matrix elements, $t_2/t_1 =0.7\approx \sqrt{J_2/J_1} $, and a value of $J_1/t_1 = 1/3$ corresponding loosely to a value of $U \approx 4t^2/J=12 t$. On the cylinders we study, the undoped system is fully gapped, so effectively corresponds to a compactified version of a gapped $Z_2$ spin liquid of the sort that arises in the quantum dimer model\cite{rokhsarandme,moessnersondhi} and the toric code model\cite{toric}. Upon lightly doping we find a state which still shows exponentially falling spin-spin correlations, with correlation lengths that are longer than but of the same order as in the undoped system.  Most importantly, we find that even at the smallest $\delta$ and on our largest 6-leg cylinders, the SC correlations are strong and decay with a slow power law, $\sim |r|^{-K_{sc}}$, with $K_{sc} \approx 1$.  This slow decay implies a SC susceptibility that diverges as $\chi_{sc} \sim T^{-(2-K_{sc})}$ as $T\to 0$. As far as we know, to date, {\it this is  the strongest indication of SC that has been found  in any DMRG study of a system on the square lattice} of width $L_y>4$.  Moreover, the SC correlations dominate over the (also apparent) CDW correlations.

{\bf Model and Method: }%
We employ DMRG\cite{White1992} to study the ground state properties of the hole-doped $t$-$J$ model on the square lattice, which is defined by the Hamiltonian%
\begin{eqnarray}\label{Eq:Ham}
H=-\sum_{ij\sigma} t_{ij} \left(\hat{c}^+_{i\sigma} \hat{c}_{j\sigma} + h.c.\right) + \sum_{ij}J_{ij}\left (\vec{S}_i\cdot \vec{S}_j - \frac{\hat{n}_i \hat{n}_j}{4} \right ). \nonumber
\end{eqnarray}
Here $\hat{c}^+_{i\sigma}$ ($\hat{c}_{i\sigma}$) is the electron creation (annihilation) operator on site $i=(x_i,y_i)$ with spin polarization $\sigma$, $\vec{S}_i$ is the spin operator and $\hat{n}_i=\sum_{\sigma}\hat{c}^+_{i\sigma}\hat{c}_{i\sigma}$ is the electron number operator. The electron hopping amplitude $t_{ij}$ is equal to $t_1$ ($t_2$) if $i$ and $j$ are NN (NNN) sites. $J_1$ and $J_2$ are the spin superexchange interactions between NN and NNN sites, respectively. The Hilbert space is constrained by the no-double occupancy condition, $n_i\leq 1$. At half-filling, i.e., $n_i=1$, 
$H$ reduces to the spin-1/2 antiferromagnetic $J_1$-$J_2$ Heisenberg model.

We take the lattice geometry to be cylindrical with periodic and open boundary conditions in the $\hat{y}$ and $\hat{x}$ directions, respectively. Here $\hat{y}=(0,1)$ and $\hat{x}=(1,0)$ are the two basis vectors of the square lattice. Here, we focus on cylinders with width $L_y$ and length $L_x$, where $L_x$ and $L_y$ are the number of sites along the $\hat{x}$ and $\hat{y}$ directions, respectively. The total number of sites is $N=L_x\times L_y$, the number of electrons $N_e$, and the doping level of the system is defined as $\delta=N_h/N$, where $N_h=N-N_e$ is the number of doped holes relative to the half-filled insulator with $N_e=N$. In the present study, we focus on  $L_y=4$ cylinders of length up to $L_x=128$ and  $L_y=6$ cylinders of length up to $L_x=48$, and for values of $\delta = 1/18$, $1/16$, and $1/12$. We set $J_1$=1 as an energy unit and $J_2=0.5$ such that the undoped system is deep in the QSL phase at half-filling.\cite{figandsondhi,Jiang2012,Gong2014,Wang2018} We consider $t_1=3$ and $t_2=t_1\ \sqrt{J_2/J_1}$ to make a connection to the corresponding Hubbard model. We perform up to 90 sweeps and keep up to $m=10000$ states for $L_y=4$ cylinders with a typical truncation error $\epsilon< 10^{-7}$, and up to $m=40000$ states for $L_y=6$ cylinders with a typical truncation error $\epsilon< 10^{-6}$. Further details of the numerical simulation are provided in the Supplemental Material (SM).

\begin{figure}
  \includegraphics[width=\linewidth]{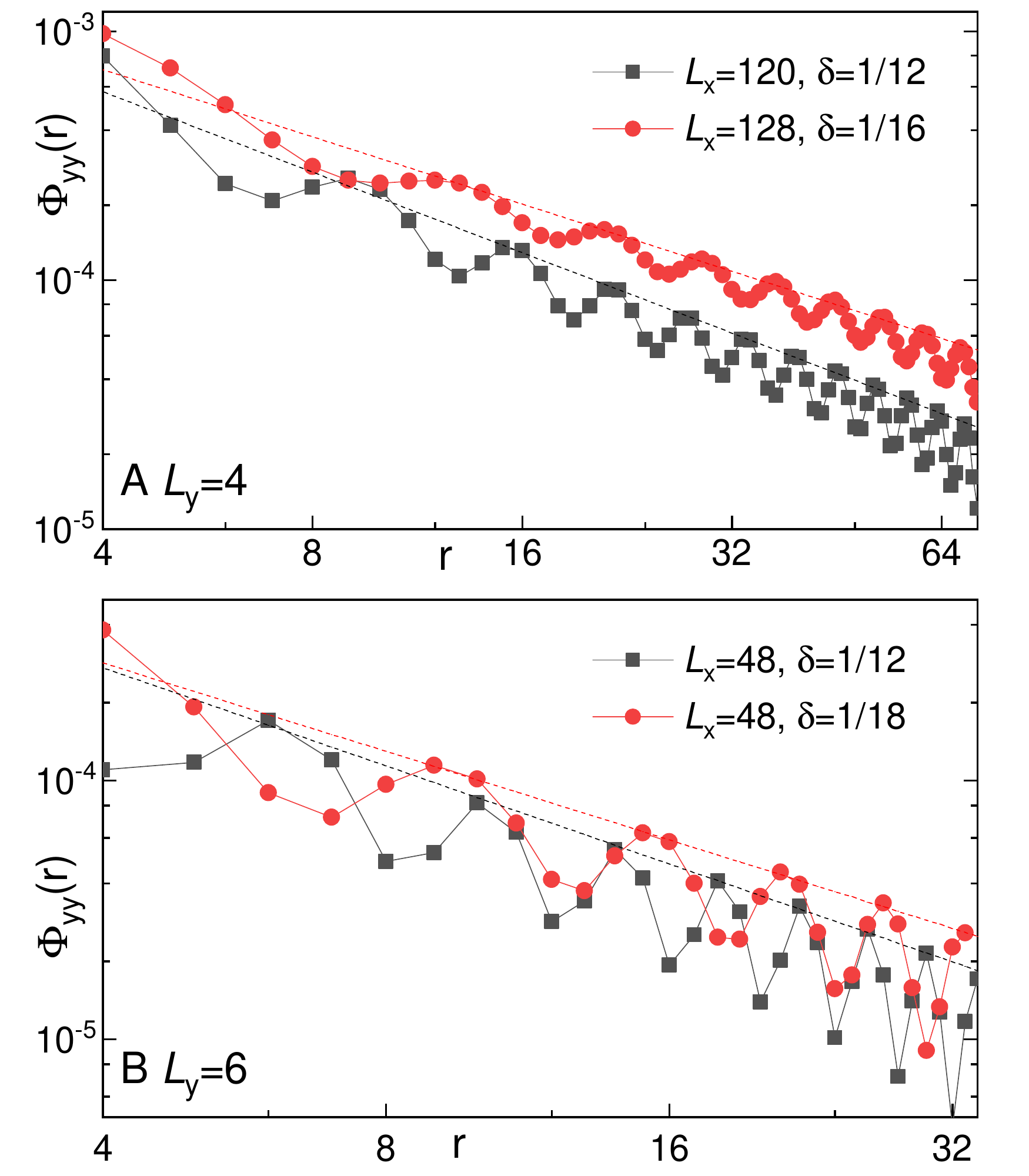}
  \caption{(Color online) Superconducting pair-field correlations $\Phi_{yy}(r)$ on  double-logarithmic scales for (A) $L_y=4$ cylinders at $\delta=1/12$ and $\delta=1/16$, and (B) $L_y=6$ cylinders at $\delta=1/12$ and $\delta=1/18$. $r$ is the distance between two Cooper pairs in the $\hat{x}$ direction. The dashed lines denote power-law fitting to $\Phi_{yy}(r)\sim r^{-K_{sc}}$.}\label{Fig:SC}
\end{figure}

{\bf Superconducting pair-field correlations: }%
We have calculated  the equal-time spin-singlet  SC pair-field correlation function %
\begin{eqnarray}
\Phi_{\alpha\beta}(r)=\frac{1}{L_y}\sum_{y=1}^{L_y}|\langle\Delta^{\dagger}_{\alpha}(x_0,y)\Delta_{\beta}(x_0+r,y)\rangle|. \label{Eq:SC}
\end{eqnarray}
$\Delta^{\dagger}_{\alpha}(x,y)=\frac{1}{\sqrt{2}}[\hat{c}^{\dagger}_{(x,y),\uparrow}\hat{c}^{\dagger}_{(x,y)+\alpha,\downarrow}+\hat{c}^{\dagger}_{(x,y)+\alpha,\uparrow}\hat{c}^{\dagger}_{(x,y),\downarrow}]$ is the spin-singlet pair creation operator on bond $\alpha=\hat{x}$ or $\hat{y}$, where ($x_0,y$) is a reference bond taken as $x_0\sim L_x/4$ and $r$ is the  displacement between bonds in the $\hat x$ direction.

Fig.\ref{Fig:SC} shows $\Phi_{yy}(r)$ for both $L_y=4$ and $L_y=6$ cylinders at different doping levels. At long distance, $\Phi(r)$ is characterized by a power-law with the appropriate Luttinger exponent $K_{sc}$ defined by%
\begin{eqnarray}
\Phi(r)\sim r^{-K_{sc}}.\label{Eq:Ksc}
\end{eqnarray}
The exponent $K_{sc}$, which is obtained by fitting the results using Eq.(\ref{Eq:Ksc}), is $K_{sc}=1.08(4)$ for $\delta=1/12$ and $K_{sc}=0.95(2)$ for $\delta=1/16$ on $L_y=4$ cylinders, and $K_{sc}=1.26(7)$ for $\delta=1/12$ and $K_{sc}=1.14(5)$ for $\delta=1/18$ on $L_y=6$ cylinders. This establishes that the lightly doped QSL on both $L_y=4$ and $L_y=6$ cylinders has quasi-long-range SC correlations. In addition to $\Phi_{yy}(r)$, we have also calculated components of the tensor -- $\Phi_{xx}(r)$ and $\Phi_{xy}(r)$ -- and find that $\Phi_{xx}(r) \sim \Phi_{yy}(r) \sim -\Phi_{xy}(r)$. In short, the SC correlations have a d-wave form.

\begin{figure}
  \includegraphics[width=\linewidth]{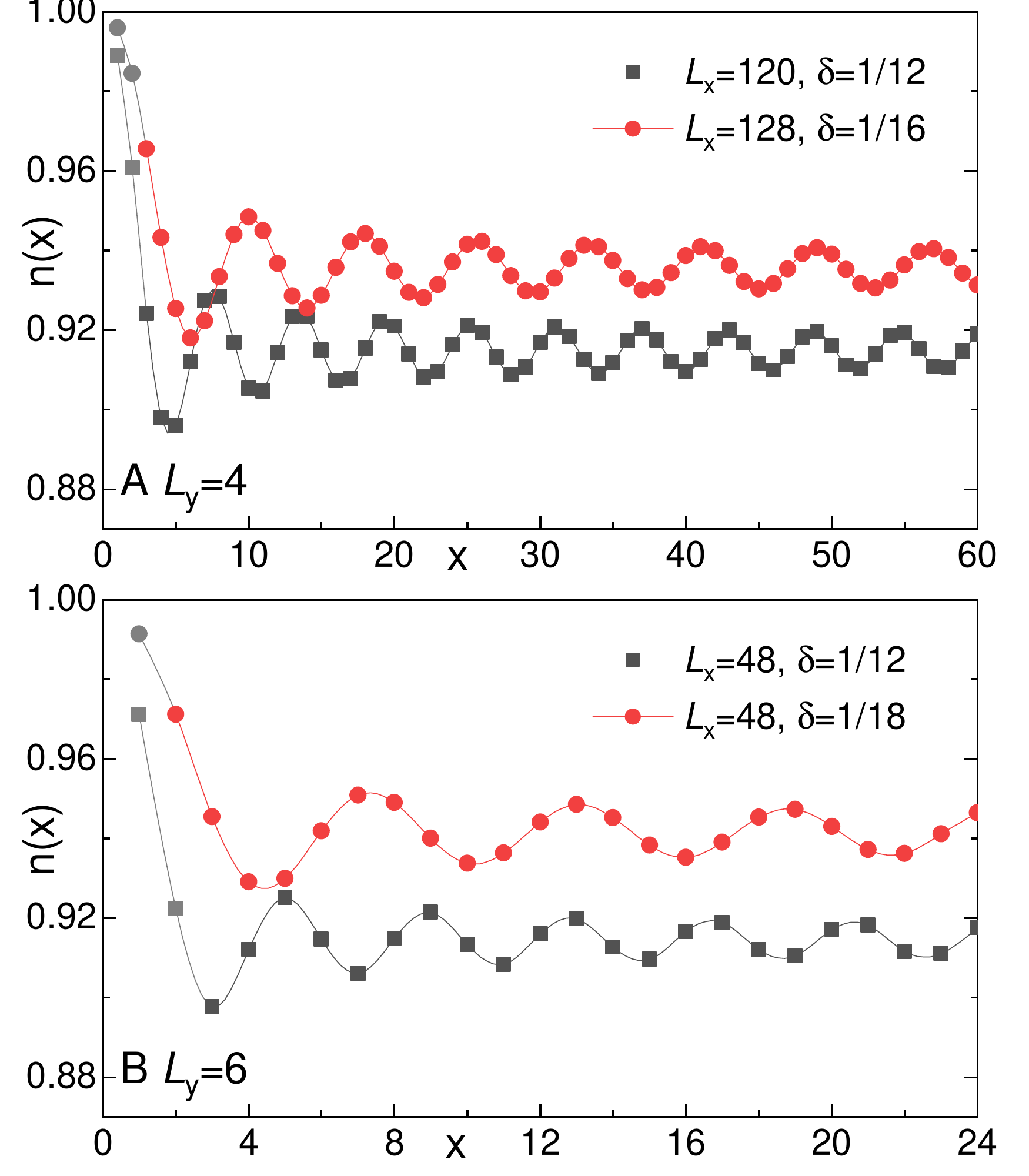}
  \caption{(Color online) Charge density profiles $n(x)$ for (A) $L_y=4$ cylinders at $\delta=1/12$ and $\delta=1/16$, and (B) $L_y=6$ cylinders at $\delta=1/12$ and $\delta=1/18$. The exponent $K_c$ is extracted using Eq.(\ref{Eq:Kc}), with the data points in grey neglected to minimize  boundary effects.}\label{Fig:CDW}
\end{figure}

{\bf CDW correlations: }%
To measure the charge order, we define the rung density operator $\hat{n}(x)=L_y^{-1}\sum_{y=1}^{L_y}\hat{n}(x,y)$ and its expectation value $n(x)=\langle \hat{n}(x)\rangle$. Fig.\ref{Fig:CDW}A shows the charge density distribution $n(x)$ for $L_y=4$ cylinders, which is consistent with ``half-filled charge stripes'' with  wavelength $\lambda=1/2\delta$. This corresponds to an ordering wavevector $Q=4\pi\delta$ corresponding to half a doped hole per 2D unit cell, i.e. viewing the cylinder as a 1D system, 2 holes per 1D unit cell. The charge density profile $n(x)$ for  $L_y=6$ cylinders is shown in Fig.\ref{Fig:CDW}B, which has wavelength $\lambda=1/3\delta$, consistent with ``third-filled" charge stripes. This corresponds to an ordering wavevector $Q=6\pi\delta$ and one third of a doped hole per 2D unit cell - again corresponding to 2 holes per 1D unit cell.

At long distance, the spatial decay of the CDW correlation is dominated by a power-law with the Luttinger exponent $K_c$. The exponent $K_c$ can be obtained by fitting the charge density oscillations (Friedel oscillations) induced by the boundaries of the cylinder\cite{White2002}
\begin{eqnarray}
n(x)=n_0 + 
A_Q\ast {\rm cos}(
Qx + \phi) x^{-K_c/2}.\label{Eq:Kc}
\end{eqnarray}
Here $A_Q$ is an amplitude, $\phi$ is a phase shift,  $n_0=1-\delta$ is the mean density, and $Q=4\pi\delta$.  Note that a few data points (Fig.\ref{Fig:CDW}A and B, light grey color) are excluded to minimize the boundary effect and improve the fitting quality. The extracted exponents for $L_y=4$ cylinders are $K_c=1.29(3)$ when $\delta=1/12$ and $K_c=1.37(3)$ when $\delta=1/16$. For $L_y=6$ cylinders, $K_c=1.42(5)$ when $\delta=1/12$ and $K_c=1.55(5)$ when $\delta=1/16$. Similarly, $K_c$ can also be obtained from the charge density-density correlation which gives consistent results (see SM).

\begin{figure}
  \includegraphics[width=\linewidth]{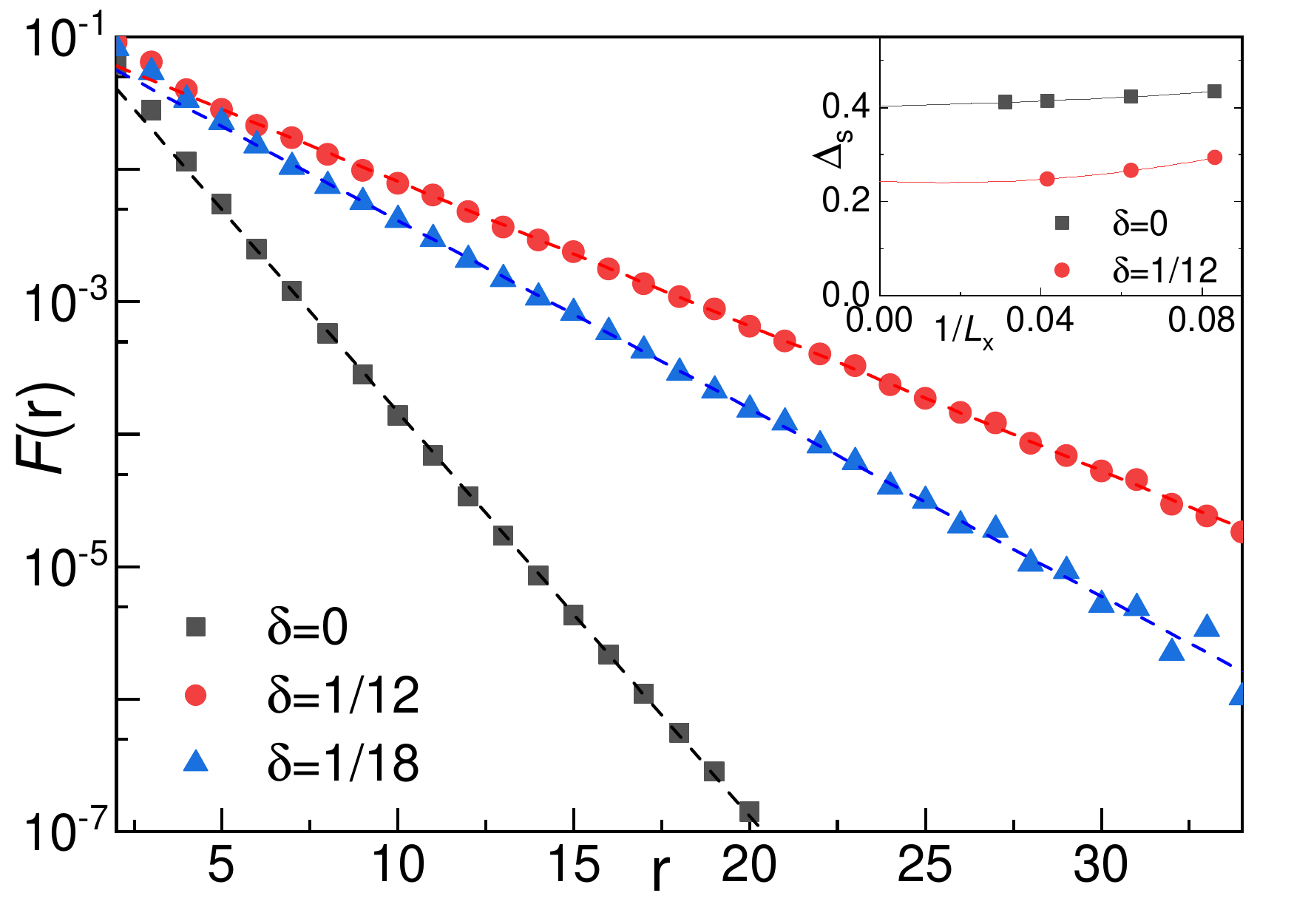}
  \caption{(Color online) Spin-spin correlations $F(r)$ for $L_y=6$ cylinders at $\delta=0$, $\delta=1/12$ and $\delta=1/18$ on the semi-logarithmic scale. Dashed lines denote exponential fit $F(r)\sim e^{-r/\xi_s}$, where $r$ is the distance between two sites in the $\hat{x}$ direction. Inset: Spin gap $\Delta_s$ for $L_y=6$ cylinders at $\delta=0$ and $\delta=1/12$. Solid lines denote second-order polynomial fitting.}\label{Fig:SpinCor}
\end{figure}

{\bf Spin-spin correlations: }%
To describe the magnetic properties of the ground state, we calculate the spin-spin correlation functions defined as%
\begin{eqnarray}\label{Eq:SpinCor}
F(r)=\frac{1}{L_y}\sum_{y=1}^{L_y}|\langle \vec{S}_{x_0,y}\cdot \vec{S}_{x_0+r,y}\rangle |. 
\end{eqnarray}
Fig.\ref{Fig:SpinCor} shows $F(r)$ for $L_y=6$ cylinders at different doping levels, which decays exponentially as $F(r)\sim e^{-r/\xi_s}$ at long-distances, with a correlation length $\xi_s=3.98(1)$ lattice spacings for $\delta=1/12$ and $\xi_s=3.06(2)$ lattice spacings for $\delta=1/18$. For comparison, the spin-spin correlation $F(r)$ at half-filling, i.e., $\delta=0$, is also shown, which decays exponentially with a correlation length $\xi_s=1.42(1)$. Therefore, the spin-spin correlations at finite doping levels are short-ranged and similar to those of the QSL at half-filling. In the inset of Fig.\ref{Fig:SpinCor}, we show the spin gap, defined as $\Delta_s=E_0(S_z=1)-E_0(S_z=0)$, where $E_0(S_z)$ is the ground state energy of a system with total spin $S_z$. At half-filling, i.e., $\delta=0$, $\Delta_s=0.40(1)$ which is consistent with previous studies.\cite{Jiang2012,Gong2014} At $\delta=1/12$, $\Delta_s=0.24(1)$, which is consistent with the short-range nature of $F(r)$.

\begin{figure}[tb]
\centering
    \includegraphics[width=1\linewidth]{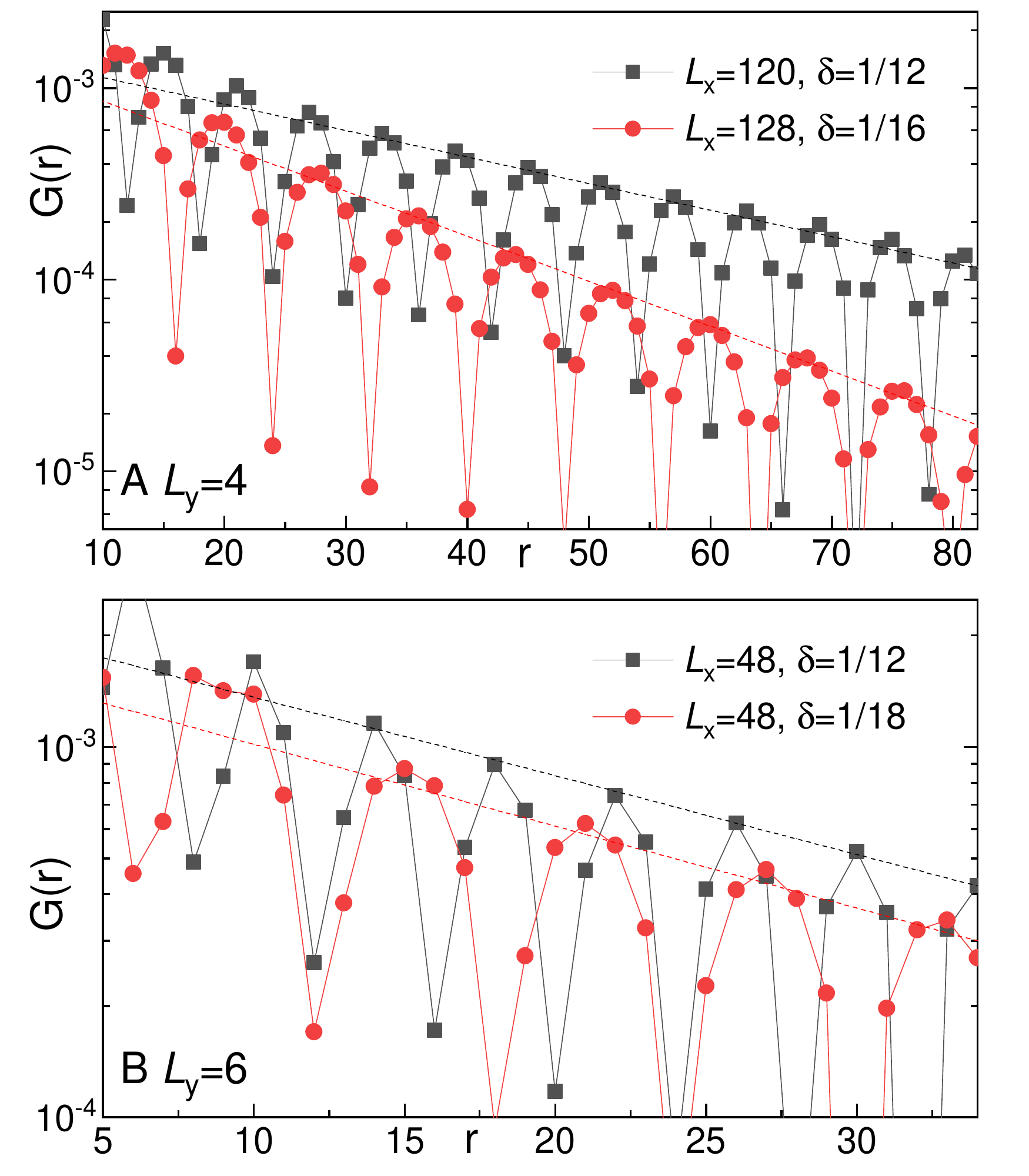}
\caption{(Color online) Single-particle Green function $G(r)$ for (A) $L_y=4$ cylinders at $\delta=1/12$ and $\delta=1/16$, and (B) $L_y=6$ cylinders at $\delta=1/12$ and $\delta=1/18$ on the semi-logarithmic scale. Dashed line denote exponential fitting $G(r)\sim e^{-r/\xi_G}$ where $r$ is the distance between two sites in the $\hat{x}$ direction.}\label{Fig:CC}
\end{figure}

\begin{figure}[htb]
\centering
    \includegraphics[width=1\linewidth]{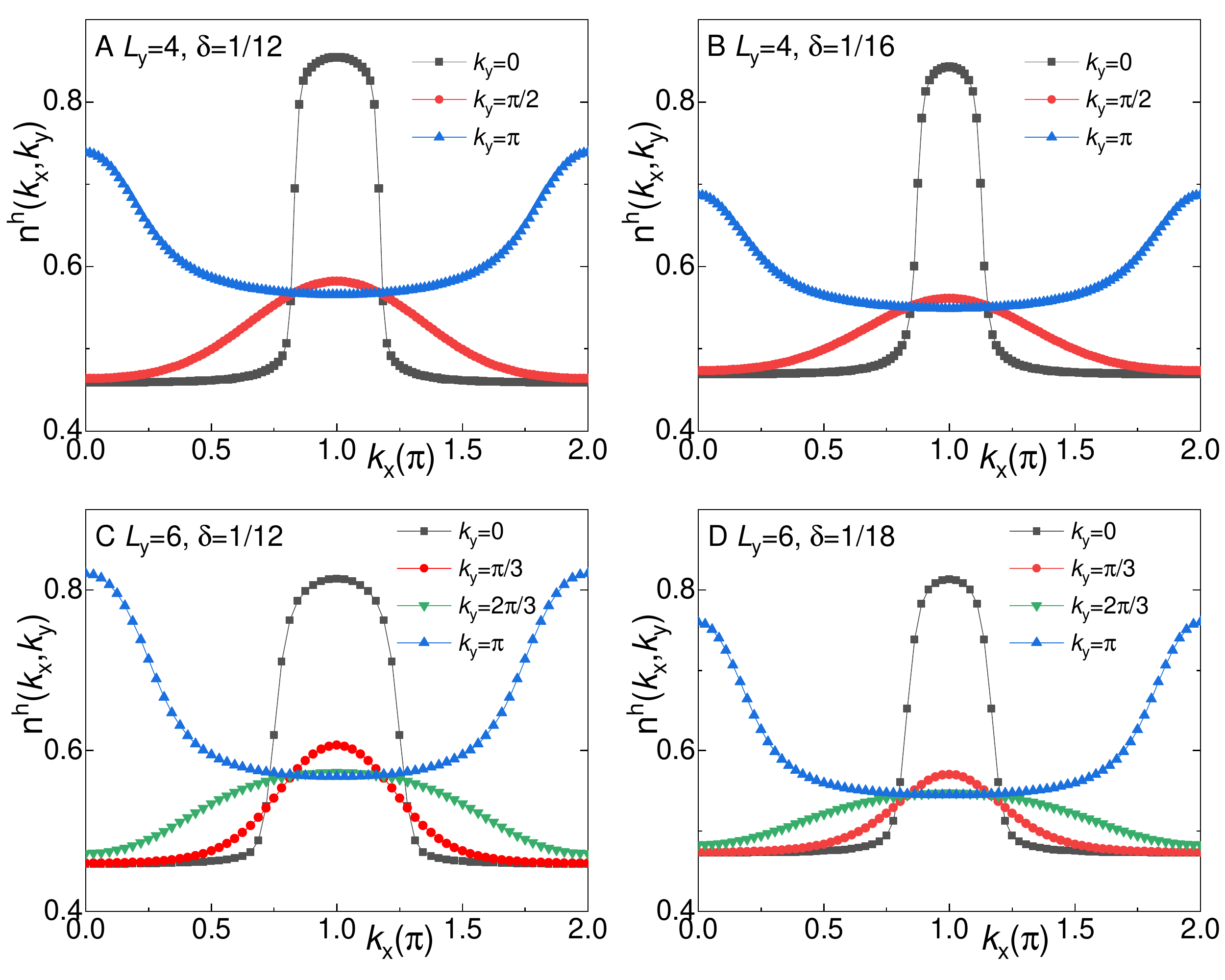}
\caption{(Color online) Hole momentum distribution function $n^h(k_x,k_y)$ for $L_y=4$ cylinders at (A) $\delta=1/12$ and (B) $\delta=1/16$, and $L_y=6$ cylinders at (C) $\delta=1/12$ and (D) $\delta=1/18$ at different $k_y$ as a function of $k_x$ in unit of $\pi$.}\label{Fig:Nk}
\end{figure}

{\bf Single particle Green function: }%
We have also calculated the single-particle Green function, defined as%
\begin{eqnarray}\label{Eq:CC}
  G(r)=\frac{1}{L_y}\sum_{y=1}^{L_y}\langle c^{\dagger}_{(x_0,y),\sigma} c_{(x_0+r,y),\sigma}\rangle.
\end{eqnarray}
Fig.\ref{Fig:CC} shows  $G(r)$ for both $L_y=4$ and $L_y=6$ cylinders at different doping levels, the long distance behavior of $G$ is consistent with exponential
decay $G(r)\sim e^{-r/\xi_G}$. The extracted correlation lengths for $L_y=4$ cylinders are $\xi_G=30(2)$ when $\delta=1/12$ and $\xi_G=18(1)$ when $\delta=1/16$, while for $L_y=6$ cylinders, $\xi_G=21(1)$ when $\delta=1/12$ and $\xi_G=20(2)$ when $\delta=1/18$.

We have also measured the hole momentum distribution function defined as%
\begin{eqnarray}\label{Eq:Nk}
n^h(\mathbf{k})=\frac{1}{2}\left[2-\sum_\sigma n_\sigma(\mathbf{k})\right].
\end{eqnarray}
Here $n_\sigma(\mathbf{k})=\frac{1}{N}\sum_{ij}e^{i\mathbf{k}\cdot (\mathbf{r}_i-\mathbf{r}_j)\langle\hat{c}^+_{i\sigma}\hat{c}_{j\sigma}}\rangle$ is the electron momentum distribution function for electron with spin-$\sigma$. Fig.\ref{Fig:Nk} shows $n^h(\mathbf{k})$ for both $L_y=4$ and $L_y=6$ cylinders at different doping levels. Not surprisingly, there are no clear discontinuities in $n^h(\mathbf{k})$ of the sort that would be expected at the Fermi momenta of a Fermi liquid.  

However,  there are sharp drops in $n^h(\mathbf{k})$ (which can be identified as maxima of $\left|dn^h(\mathbf{k})/dk_x\right|$) (See SM for details) that are suggestive of the ``near existence'' of a  Fermi surface.  These features are most prominent for $k_y=0$, where they occur at $k_x \approx \pi \pm k_0$, but there are slightly broader features of the same general sort for $k_y=\pi$,  at $k_x =  \pm k_\pi$.  For $L_y=4$ and  $\delta=1/16$, $k_0=0.075\pi$ and $k_\pi=0.175\pi$;  for $L_y=4$ and $\delta=1/12$, $k_0=0.175\pi$ and $k_\pi=0.192\pi$; for $L_y=6$ and  $\delta=1/18$, $k_0=0.17\pi$ and $k_\pi=0.18\pi$; for $L_y=6$ and  $\delta=1/12$, $k_0=0.25\pi$ and $k_\pi=0.25\pi$.
Within the numerical uncertainty, there is a direct relation between these quasi-Fermi momenta and the CDW ordering vector: $Q= 2(k_0+k_\pi)$.  Moreover, since $Q=\pi L_y \delta$, this corresponds to the expected value of $2k_F$ that would correspond to the ``volume'' of the Fermi surface under conditions (not satisfied in the present case) in which  Luttinger's theorem applies.

{\bf Conclusion:} There is necessarily  a speculative leap from  results on finite cylinders to the 2D limit. However, we feel 
that the present results - and those of a similar study by one of us on the  triangular lattice $t$-$J$ model on 4 and 6 leg cylinders\cite{hongchentriangle,Jiang2020} - can plausibly be taken as representative of the solution of the corresponding 2D problem.  In particular, they support the proposition that SC can emerge upon light doping of a QSL. Conversely, our earlier observation of an insulating holon crystal in a lightly doped Kagome system\cite{Jiang2017,Peng2021} and CDW order in a lightly doped honeycomb Kiatev spin liquid\cite{Peng2020} imply that SC is not the universal result of doping a QSL. Indeed, for otherwise identical cylinders to those reported above, reversing the sign of $t_2$ (i.e. taking $t_2 = - t_1\sqrt{J_2/J_1}$) reduces the long distance SC correlations by many orders of magnitude although whether some weak SC power-law correlations persist is still unsettled. Moreover, we have also found greatly enhanced SC correlations on 4 and 6 leg cylinders with  a spatially modulated (``striped'') version of the square-lattice Hubbard model\cite{stripeHubbard};   it thus may be aspects of doping a quantum paramagnet (i.e. a system in which quantum fluctuations are sufficient to destroy magnetic order) rather than specific features of a doped QSL that is  responsible for the strong SC tendencies.

It is harder still to make inferences about $T_c$ itself in the 2D limit.  The large values of the spin-gaps, $\Delta_s \sim J/4$,  are suggestive that pairing is sufficiently strong to persist to very high $T$.  It is therefore likely that $T_c$ is determined by the phase ordering scale,\cite{emeryphase} in other words that the zero temperature superfluid stiffness and hence $T_c$ itself rise roughly linearly with $\delta$ for $\delta\ll 1$.\cite{KRS} 

{\it Acknowledgments:} We would like to thank Erez Berg, Johannes Motruk, Zheng-Yu Weng and Yi-Fan Jiang for helpful discussions. This work was supported by the Department of Energy, Office of Science, Basic Energy Sciences, Materials Sciences and Engineering Division, under Contract DE-AC02-76SF00515.


\begin{thebibliography}{10}

\bibitem{anderson87}
P.~W. Anderson.
\newblock {The Resonating valence bond state in La$_2$CuO$_4$ and
  superconductivity}.
\newblock {\em Science}, 235(4793):1196--1198, 1987.

\bibitem{emery87}
V.~J. Emery.
\newblock {Theory of high-${\mathrm{T}}_{\mathrm{c}}$ superconductivity in
  oxides}.
\newblock {\em Phys. Rev. Lett.}, 58:2794--2797, Jun 1987.

\bibitem{scalhubbard}
D.~J. {Scalapino}.
\newblock {A common thread: The pairing interaction for unconventional
  superconductors}.
\newblock {\em Reviews of Modern Physics}, 84(4):1383--1417, October 2012.

\bibitem{complexity}
Sophia Kivelson and Steven Kivelson.
\newblock {Understanding complexity}.
\newblock {\em {Nature Physics}}, {14}({5}):{426--427}, {May} {2018}.

\bibitem{Raghu2010}
S.~Raghu, S.~A. Kivelson, and D.~J. Scalapino.
\newblock {Superconductivity in the repulsive Hubbard model: An asymptotically
  exact weak-coupling solution}.
\newblock {\em Phys. Rev. B}, 81:224505, Jun 2010.

\bibitem{gull}
E.~Gull and A.~J. Millis.
\newblock Quasiparticle properties of the superconducting state of the
  two-dimensional hubbard model.
\newblock {\em Phys. Rev. B}, 91:085116, Feb 2015.

\bibitem{maier}
P.~Staar, M.~Jiang, U.~R. H\"ahner, T.~C. Schulthess, and T.~A. Maier.
\newblock Interlaced coarse-graining for the dynamic cluster approximation.
\newblock {\em Phys. Rev. B}, 93:165144, Apr 2016.

\bibitem{andrey}
Yuxuan Wang, Artem Abanov, Boris~L. Altshuler, Emil~A. Yuzbashyan, and
  Andrey~V. Chubukov.
\newblock Superconductivity near a quantum-critical point: The special role of
  the first matsubara frequency.
\newblock {\em Phys. Rev. Lett.}, 117:157001, Oct 2016.

\bibitem{Jiang2018tJ}
Hong-Chen Jiang, Zheng-Yu Weng, and Steven~A. Kivelson.
\newblock Superconductivity in the doped $\mathit{t}\ensuremath{-}\mathit{J}$
  model: Results for four-leg cylinders.
\newblock {\em Phys. Rev. B}, 98:140505, Oct 2018.

\bibitem{Jiang2019Hub}
Hong-Chen Jiang and Thomas~P. Devereaux.
\newblock Superconductivity in the doped hubbard model and its interplay with
  next-nearest hopping t'.
\newblock {\em Science}, 365(6460):1424--1428, 2019.

\bibitem{Jiang2020Hub}
Yi-Fan Jiang, Jan Zaanen, Thomas~P. Devereaux, and Hong-Chen Jiang.
\newblock Ground state phase diagram of the doped hubbard model on the four-leg
  cylinder.
\newblock {\em Phys. Rev. Research}, 2:033073, Jul 2020.

\bibitem{Jiang2020tJ}
Hong-Chen Jiang, Shuai Chen, and Zheng-Yu Weng.
\newblock Critical role of the sign structure in the doped mott insulator:
  Luther-emery versus fermi-liquid-like state in quasi-one-dimensional ladders.
\newblock {\em Phys. Rev. B}, 102:104512, Sep 2020.

\bibitem{Chung2020}
Chia-Min Chung, Mingpu Qin, Shiwei Zhang, Ulrich Schollw\"ock, and Steven~R.
  White.
\newblock Plaquette versus ordinary $d$-wave pairing in the
  ${t}^{\ensuremath{'}}$-hubbard model on a width-4 cylinder.
\newblock {\em Phys. Rev. B}, 102:041106, Jul 2020.

\bibitem{Simons2015}
J.~P.~F. LeBlanc, Andrey~E. Antipov, Federico Becca, Ireneusz~W. Bulik, Garnet
  Kin-Lic Chan, Chia-Min Chung, Youjin Deng, Michel Ferrero, Thomas~M.
  Henderson, Carlos~A. Jim\'enez-Hoyos, E.~Kozik, Xuan-Wen Liu, Andrew~J.
  Millis, N.~V. Prokof'ev, Mingpu Qin, Gustavo~E. Scuseria, Hao Shi, B.~V.
  Svistunov, Luca~F. Tocchio, I.~S. Tupitsyn, Steven~R. White, Shiwei Zhang,
  Bo-Xiao Zheng, Zhenyue Zhu, and Emanuel Gull.
\newblock {Solutions of the two-dimensional Hubbard model: benchmarks and
  results from a wide range of numerical algorithms}.
\newblock {\em Phys. Rev. X}, 5:041041, Dec 2015.

\bibitem{whitenew}
Chia-Min Chung, Mingpu Qin, Shiwei Zhang, Ulrich Schollw\"ock, and Steven~R.
  White.
\newblock {Plaquette versus ordinary $d$-wave pairing in the $t'$-Hubbard model
  on a width-4 cylinder}.
\newblock {\em Phys. Rev. B}, 102:041106, Jul 2020.

\bibitem{sorella}
Sandro {Sorella}.
\newblock {The phase diagram of the Hubbard model by variational auxiliary
  field quantum Monte Carlo}.
\newblock {\em arXiv e-print}, page arXiv:2101.07045, Jan 2021.

\bibitem{emeryandlin}
VJ~{Emery}, HQ~{Lin}, and SA~{Kivelson}.
\newblock {Phase separation in the t-J model}.
\newblock {\em \prl}, 64:475--478, Jan 1990.

\bibitem{zaanenstripes}
Jan Zaanen and Olle Gunnarsson.
\newblock {Charged magnetic domain lines and the magnetism of high-${T}_{\rm
  c}$ oxides}.
\newblock {\em Phys. Rev. B}, 40:7391--7394, Oct 1989.

\bibitem{shultzstripes1}
{Schulz, H.J.}
\newblock {Domain walls in a doped antiferromagnet}.
\newblock {\em J. Phys. France}, 50(18):2833--2849, 1989.

\bibitem{machidastripes1}
Kazushige Machida.
\newblock {Magnetism in La$_2$CuO$_4$ based compounds}.
\newblock {\em Physica C: Superconductivity}, 158(1):192 -- 196, 1989.

\bibitem{whitescalapino4leg}
S.~R. White and D.~J. Scalapino.
\newblock {Ground states of the doped four-leg $t$-$J$ ladder}.
\newblock {\em {Phys. Rev. B}}, {55}({22}):{14701--14704}, {June} {1997}.

\bibitem{Qin2020}
Mingpu Qin, Chia-Min Chung, Hao Shi, Ettore Vitali, Claudius Hubig, Ulrich
  Schollw\"ock, Steven~R. White, and Shiwei Zhang.
\newblock {Absence of superconductivity in the pure two-dimensional Hubbard
  model}.
\newblock {\em Phys. Rev. X}, 10:031016, Jul 2020.

\bibitem{KRS}
SA~Kivelson, DS~Rokhsar, and JP~Sethna.
\newblock {Topology of the resonating valence-bond state - solitons and High
  T$_c$ superconductivity}.
\newblock {\em {Phys. Rev. B}}, {35}({16}):{8865--8868}, {June 1} {1987}.

\bibitem{laughlinsc}
R.~B. {Laughlin}.
\newblock {The relationship between high-temperature superconductivity and the
  fractional quantum Hall effect}.
\newblock {\em Science}, 242:525--533, October 1988.

\bibitem{kotliar}
G~Kotliar.
\newblock {Resonating valence bonds and d-wave superconductivity}.
\newblock {\em {Phys. Rev. B }}, {37}({7}):{3664--3666}, {MAR 1} {1988}.

\bibitem{balentsandnayak}
L.~Balents, M.~P.~A. Fisher, and C.~Nayak.
\newblock {Nodal liquid theory of the pseudo-gap phase of high-$T_{\rm c}$
  superconductors}.
\newblock {\em {Int. Jour. Mod. Phys. B}}, {12}({10}):{1033--1068}, {Apr}
  {1998}.

\bibitem{vanilla}
PW~Anderson, PA~Lee, M~Randeria, TM~Rice, N~Trivedi, and FC~Zhang.
\newblock {The physics behind high-temperature superconducting cuprates: the
  `plain vanilla' version of RVB}.
\newblock {\em {Journal of Physics-Condensed Matter}}, {16}({24}):{R755--R769},
  {June 23} {2004}.

\bibitem{Song2021}
Xue-Yang Song, Ashvin Vishwanath, and Ya-Hui Zhang.
\newblock Doping the chiral spin liquid -- topological superconductor or chiral
  metal?
\newblock arXiv:2011.10044.

\bibitem{figandsondhi}
F.~Figueirido, A.~Karlhede, S.~Kivelson, S.~Sondhi, M.~Rocek, and D.~S.
  Rokhsar.
\newblock {Exact diagonalization of finite frustrated spin-$\frac{1}{2}$
  Heisenberg models}.
\newblock {\em Phys. Rev. B}, 41:4619--4632, Mar 1990.

\bibitem{Capriotti2001}
Luca Capriotti, Federico Becca, Alberto Parola, and Sandro Sorella.
\newblock Resonating valence bond wave functions for strongly frustrated spin
  systems.
\newblock {\em Phys. Rev. Lett.}, 87:097201, Aug 2001.

\bibitem{Jiang2012}
Hong-Chen Jiang, Hong Yao, and Leon Balents.
\newblock Spin liquid ground state of the spin-$\frac{1}{2}$ square
  ${J}_{1}$-${J}_{2}$ heisenberg model.
\newblock {\em Phys. Rev. B}, 86:024424, Jul 2012.

\bibitem{Hu2013}
Wen-Jun Hu, Federico Becca, Alberto Parola, and Sandro Sorella.
\newblock Direct evidence for a gapless ${Z}_{2}$ spin liquid by frustrating
  n\'eel antiferromagnetism.
\newblock {\em Phys. Rev. B}, 88:060402, Aug 2013.

\bibitem{Gong2014}
Shou-Shu Gong, Wei Zhu, D.~N. Sheng, Olexei~I. Motrunich, and Matthew P.~A.
  Fisher.
\newblock Plaquette ordered phase and quantum phase diagram in the
  spin-$\frac{1}{2}$ ${J}_{1}\text{\ensuremath{-}}{J}_{2}$ square heisenberg
  model.
\newblock {\em Phys. Rev. Lett.}, 113:027201, Jul 2014.

\bibitem{Morita2015}
Satoshi Morita, Ryui Kaneko, and Masatoshi Imada.
\newblock Quantum spin liquid in spin 1/2 j1-j2 heisenberg model on square
  lattice: Many-variable variational monte carlo study combined with
  quantum-number projections.
\newblock {\em Journal of the Physical Society of Japan}, 84(2):024720, 2015.

\bibitem{Wang2016}
Ling Wang, Zheng-Cheng Gu, Frank Verstraete, and Xiao-Gang Wen.
\newblock Tensor-product state approach to spin-$\frac{1}{2}$ square
  ${J}_{1}\text{\ensuremath{-}}{J}_{2}$ antiferromagnetic heisenberg model:
  Evidence for deconfined quantum criticality.
\newblock {\em Phys. Rev. B}, 94:075143, Aug 2016.

\bibitem{Wang2018}
Ling Wang and Anders~W. Sandvik.
\newblock Critical level crossings and gapless spin liquid in the
  square-lattice spin-$1/2$ ${J}_{1}\ensuremath{-}{J}_{2}$ heisenberg
  antiferromagnet.
\newblock {\em Phys. Rev. Lett.}, 121:107202, Sep 2018.

\bibitem{rokhsarandme}
Daniel~S. Rokhsar and Steven~A. Kivelson.
\newblock {Superconductivity and the Quantum Hard-Core Dimer Gas}.
\newblock {\em Phys. Rev. Lett.}, 61:2376--2379, Nov 1988.

\bibitem{moessnersondhi}
R.~Moessner and S.~L. Sondhi.
\newblock {Resonating valence bond phase in the triangular lattice quantum
  dimer model}.
\newblock {\em Phys. Rev. Lett.}, 86:1881--1884, Feb 2001.

\bibitem{toric}
AY~Kitaev.
\newblock {Fault-tolerant quantum computation by anyons}.
\newblock {\em {Annals of Physics}}, {303}({1}):{2--30}, {Jan} {2003}.

\bibitem{White1992}
Steven~R. White.
\newblock Density matrix formulation for quantum renormalization groups.
\newblock {\em Phys. Rev. Lett.}, 69:2863--2866, Nov 1992.

\bibitem{White2002}
Steven~R. White, Ian Affleck, and Douglas~J. Scalapino.
\newblock Friedel oscillations and charge density waves in chains and ladders.
\newblock {\em Phys. Rev. B}, 65:165122, Apr 2002.

\bibitem{hongchentriangle}
Hong-Chen {Jiang}.
\newblock {Superconductivity in the doped quantum spin liquid on the triangular
  lattice}.
\newblock {\em arXiv:1912.06624}, December 2019.

\bibitem{Jiang2020}
Yi-Fan Jiang and Hong-Chen Jiang.
\newblock Topological superconductivity in the doped chiral spin liquid on the
  triangular lattice.
\newblock {\em Phys. Rev. Lett.}, 125:157002, Oct 2020.

\bibitem{Jiang2017}
Hong-Chen Jiang, T.~Devereaux, and S.~A. Kivelson.
\newblock {Holon Wigner crystal in a lightly doped kagome quantum spin liquid}.
\newblock {\em Phys. Rev. Lett.}, 119:067002, Aug 2017.

\bibitem{Peng2021}
Cheng Peng, Yi-Fan Jiang, Dong-Ning Sheng, and Hong-Chen Jiang.
\newblock Doping quantum spin liquids on the kagome lattice.
\newblock {\em Advanced Quantum Technologies}, 4(3):2000126, 2021.

\bibitem{Peng2020}
Cheng {Peng}, Yi-Fan {Jiang}, Thomas~P. {Devereaux}, and Hong-Chen {Jiang}.
\newblock {Evidence of pair-density wave in doping Kitaev spin liquid on the
  honeycomb lattice}.
\newblock August 2020.

\bibitem{stripeHubbard}
Hong-Chen Jiang and Steven~A. Kivelson.
\newblock Stripe order enhanced superconductivity in the hubbard model.
\newblock {\em unpublished}, 2021.

\bibitem{emeryphase}
VJ~Emery and SA~Kivelson.
\newblock {Importance of phase fluctuations in superconductors with small
  superfluid density}.
\newblock {\em {Nature}}, {374}({6521}):{434--437}, {Mar 30} {1995}.

\bibitem{Calabrese2004}
P.~Calabrese and J.~Cardy.
\newblock {Entanglement entropy and quantum field theory}.
\newblock {\em J. Stat. Mech. Theory Exp.}, 2004(6), 2004.

\bibitem{Fagotti2011}
M.~Fagotti and P.~Calabrese.
\newblock {Universal parity effects in the entanglement entropy of XX chains
  with open boundary conditions}.
\newblock {\em J. Stat. Mech. Theory Exp.}, 2011(1), Jan 2011.

\end{thebibliography}

\clearpage
\newpage

\renewcommand{\thefigure}{S\arabic{figure}}
\setcounter{figure}{0}
\renewcommand{\theequation}{S\arabic{equation}}
\setcounter{equation}{0}
\setcounter{page}{1}
\setcounter{table}{0}

\begin{center}
\noindent {\large {\bf Supplemental Material}}
\end{center}

\section{More numerical details}\label{SM:Detail} %
We have checked the numerical convergence of our DMRG simulations regarding various symmetries such as spin rotational symmetry. It is known that the ground state of finite systems cannot spontaneously break continuous symmetry. Therefore, the true ground state of the $t$-$J$ model on finite cylinders should preserve the $SU(2)$ spin rotational symmetry. This can be considered as one of the key signatures to determine whether a DMRG simulation has converged to the true ground state. We take two routes to address this issue in our DMRG simulation. First of all, we calculate the expectation value of the $z$-component spin operator $\langle\hat{S}^z_i\rangle$, which should be zero on any lattice site $i$ in the true ground state. Indeed we find that $\langle \hat{S}^z_i\rangle=0$ on all site $i$ for both $L_y=4$ and $L_y=6$ cylinders even for the smallest kept number of states $m=2187$, which suggests that our simulations have converged. Similarly, the spin $SU(2)$ symmetry requires that the relation $\langle S^x_i S^x_j\rangle$=$\langle S^y_i S^y_j\rangle$=$\langle S^z_i S^z_j\rangle$ holds between two arbitrary sites $i$ and $j$, which is again fulfilled in our simulations. In addition to spin rotational symmetry, other symmetries including both the lattice translational symmetry in the $\hat{y}$ direction and reflection symmetry in the $\hat{x}$ direction are also satisfied. As further tests, we have also explored the effect of cylinder size, boundary effects and the pinning field. This gives us the same results as we start from a completely random initial state, which further demonstrates the reliability of our study.

\begin{figure}
  \includegraphics[width=\linewidth]{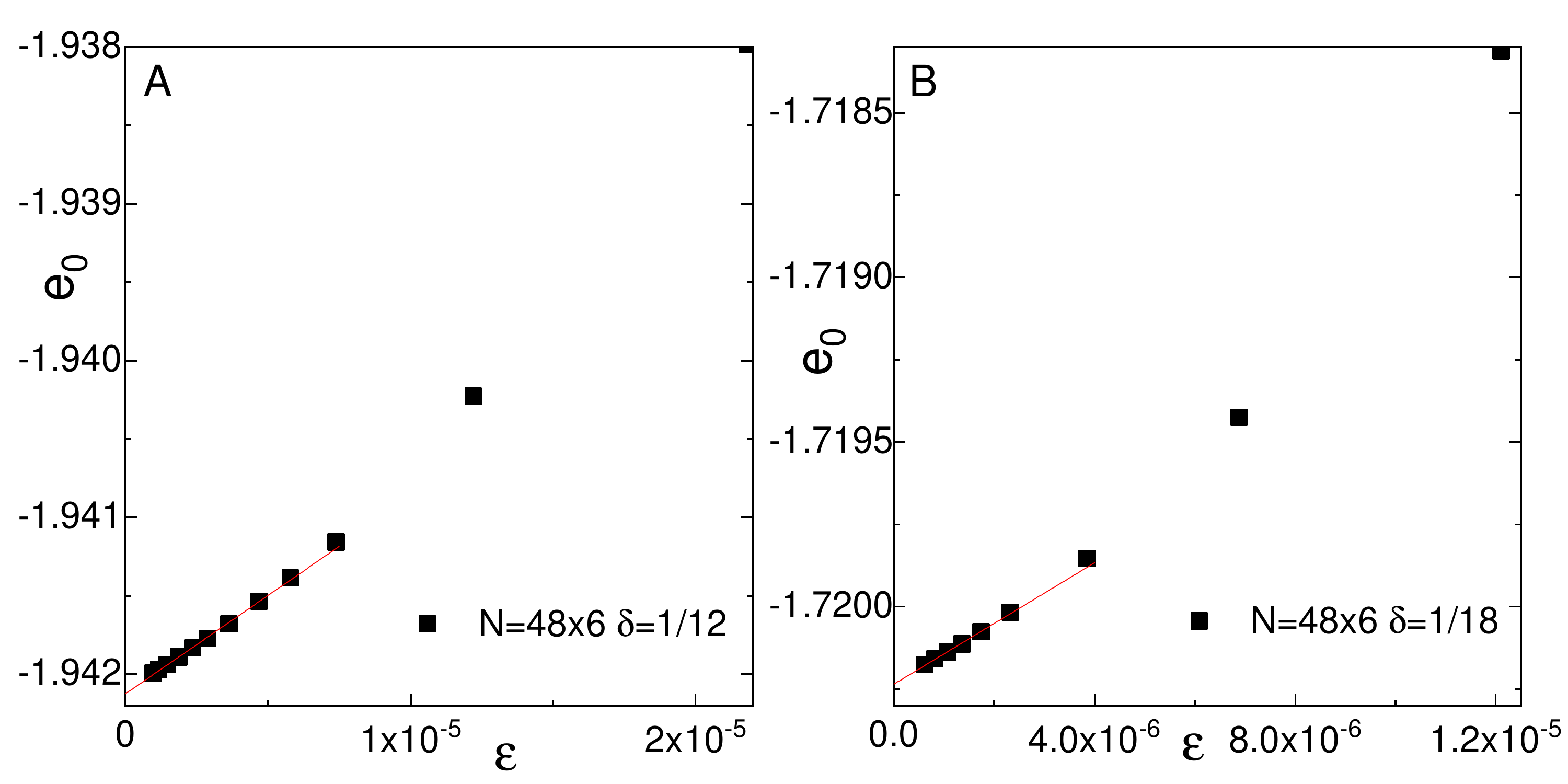}
  \caption{Ground state energy of the $t$-$J$ model. Ground state energy per site $e_0$ for (A) $\delta=1/12$ and (B) $\delta=1/18$ on $N=48\times 6$ cylinders, as a function of truncation error $\epsilon$.The red lines show the extrapolation using a linear function.}\label{FigS:EG}
\end{figure}

\section{Ground state energy} %
Fig.\ref{FigS:EG} shows examples of truncation error $\epsilon$ extrapolation of the ground state energy per site $e_0=E_0/N$, where $E_0$ is the total ground state energy and $N$ is the size of the system. For $N=48\times 6$ cylinders, by keeping $m=2187\sim 40000$ 
states for $\delta=1/12$, and $m=2187\sim 30000$ 
states for $\delta=1/18$, we are able to converge to the true ground state of the system. The truncation error extrapolation using a linear function with $m=10000\sim 40000$ for $\delta=1/12$ gives $e_0=-1.94212(1)$. For $\delta=1/18$, the truncation error extrapolation using a linear function with $m=12000\sim 30000$ gives $e_0=-1.72024(1)$. The ground state energy $e_0$ for other cylinders and doping concentrations $\delta$ can be obtained similarly. For instance, we get $e_0=-1.99745(1)$ for $N=72\times 4$ cylinder at $\delta=1/12$, and $e_0=-1.82049(1)$ for $N=96\times 4$ cylinder at $\delta=1/16$.

\begin{figure}
  \includegraphics[width=\linewidth]{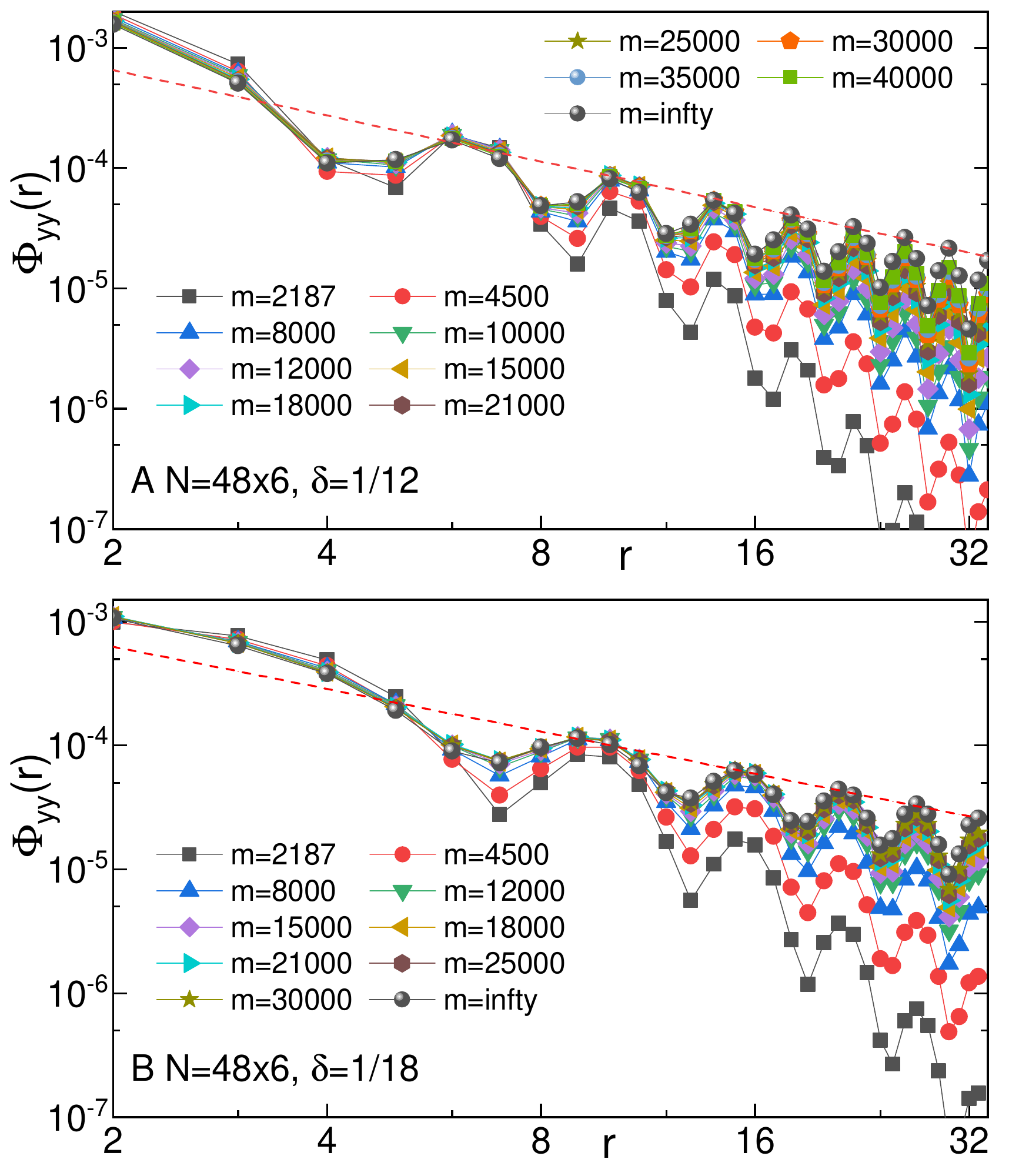}
  \caption{Convergence of SC correlations. SC correlation $\Phi_{yy}(r)$ on (A) $N=48\times 6$ cylinder at $\delta=1/12$ and (B) $N=48\times 6$ 
  cylinder at $\delta=1/18$, by keeping $m$ number of states and its extrapolation in the limit $m=\infty$. All figures are plotted 
  on double-logarithmic scales, where $r$ is the distance between two Cooper pairs in the $\mathbf{e}_x$ direction. The red dashed lines label the power-law fit $\Phi_{yy}(r)\sim 1/r^{K_{sc}}$.}\label{FigS:SC}
\end{figure}

\section{Superconducting correlations }\label{SM:SC} %
Fig.\ref{FigS:SC} shows the superconducting (SC) pair-field correlations $\Phi_{yy}(r)$ for $N=48\times 6$ cylinders at $\delta=1/12$ and $\delta=1/18$.
The extrapolated $\Phi_{yy}(r)$ in the limit $m=\infty$ or $\epsilon=0$ is obtained using second-order polynomial function with four data points of largest number of states. Here $r$ is the distance between two Cooper pairs in the $\hat{x}$ direction. To minimize the boundary and finite-size effect, the first few data points with small $r$ are excluded. As indicated by the red dashed lines, the SC correlations are consistent with a power-law decay $\Phi_{yy}(r)\propto r^{-K_{sc}}$ with $K_{sc}\sim 1$. We have used the same procedure for $L_y=4$ cylinders and obtained the extrapolated $\Phi_{yy}(r)$ in the limit $m=\infty$ or $\epsilon=0$, for which the exponent $K_{sc}
\sim 1$ can be obtained accordingly. In addition to the spin-singlet SC correlation, we have also calculated the spin-triplet SC correlation. However, it is much weaker than the spin-singlet SC correlation, suggesting that spin-triplet superconductivity is unlikely.

\begin{figure}
  \includegraphics[width=\linewidth]{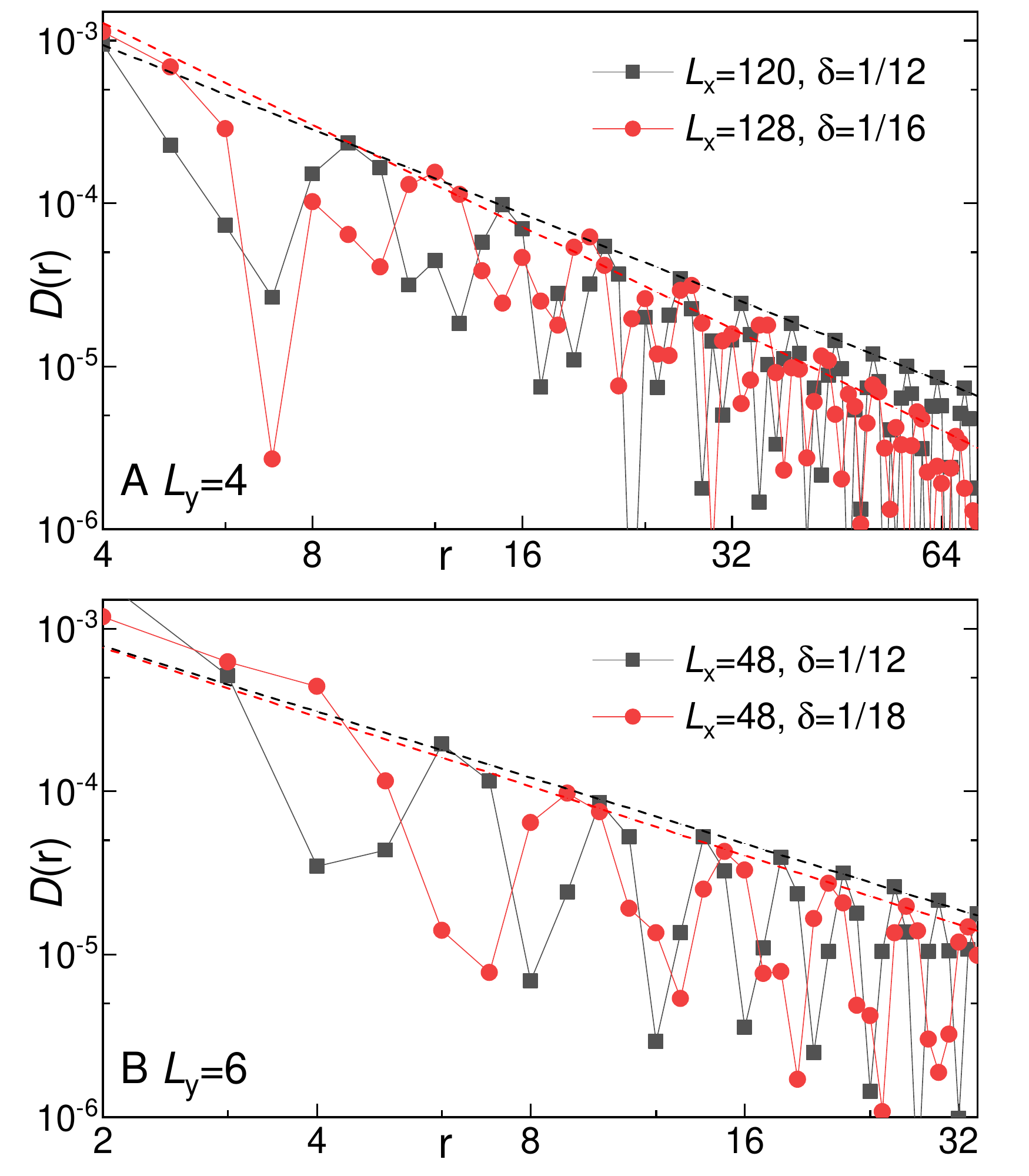}
  \caption{Charge density-density correlations. Charge density-density correlations $D(r)$ for (A) $L_y=4$ and (B) $L_y=6$ cylinders at different doping levels $\delta$ on double-logarithmic scales, where $r$ is the distance between two sites in the $\hat{x}$ direction. The dashed lines denote a power-law fit $D(r)\sim r^{-K_c}$.}\label{FigS:DenCor}
\end{figure}

\section{Charge density-density correlations}\label{SM:DenCor} %
In addition to the charge density oscillation (Friedel oscillation), the exponent $K_c$ can also be extracted from the charge density-density correlation, which is defined as $D(r)=\langle (\hat{n}(x_0)-\langle \hat{n}(x_0)\rangle)(\hat{n}(x_0+r)-\langle \hat{n}(x_0+r)\rangle)\rangle$. Here $x_0$ is the rung index of the reference site and $r$ is the distance between two sites in the $\hat{x}$ direction. Following similar procedure as for $n(x)$ and $\Phi(r)$, the extrapolated $D(r)$ for a given cylinder in the limit $\epsilon=0$ or $m=\infty$ is obtained using second-order polynomial function with four data points of largest number of states.

As shown in Fig.\ref{FigS:DenCor}A and B for both $L_y=4$ and $L_y=6$ cylinders, $D(r)$ decays with a power-law at long distances, whose exponent $K_c$ was obtained by fitting the results using $D(r)\propto r^{-K_c}$. The extracted exponents for $L_y=4$ cylinders are $K_c=1.7(1)$ when $\delta=1/12$ and $K_c=2.1(1)$ when $\delta=1/16$. For $L_y=6$ cylinders, $K_c=1.35(5)$ when $\delta=1/12$ and $K_c=1.41(4)$ when $\delta=1/16$. For comparison, the values already quoted in the main text extracted from the charge density oscillations in $n(x)$ are $K_c=1.29(3)$ when $\delta=1/12$ and $K_c=1.37(3)$ when $\delta=1/16$ for $L_y=4$ cylinders. For $L_y=6$ cylinders, $K_c=1.42(5)$ when $\delta=1/12$ and $K_c=1.55(5)$ when $\delta=1/16$. Note that $K_c$ extracted from $D(r)$ is slightly different from that extracted from the charge density oscillation $n(x)$, which may be attributable to the fact that the calculation of $D(r)$ is less accurate than $n(x)$ in the DMRG simulation. However, they are qualitatively consistent with each other.

\begin{figure}
  \includegraphics[width=\linewidth]{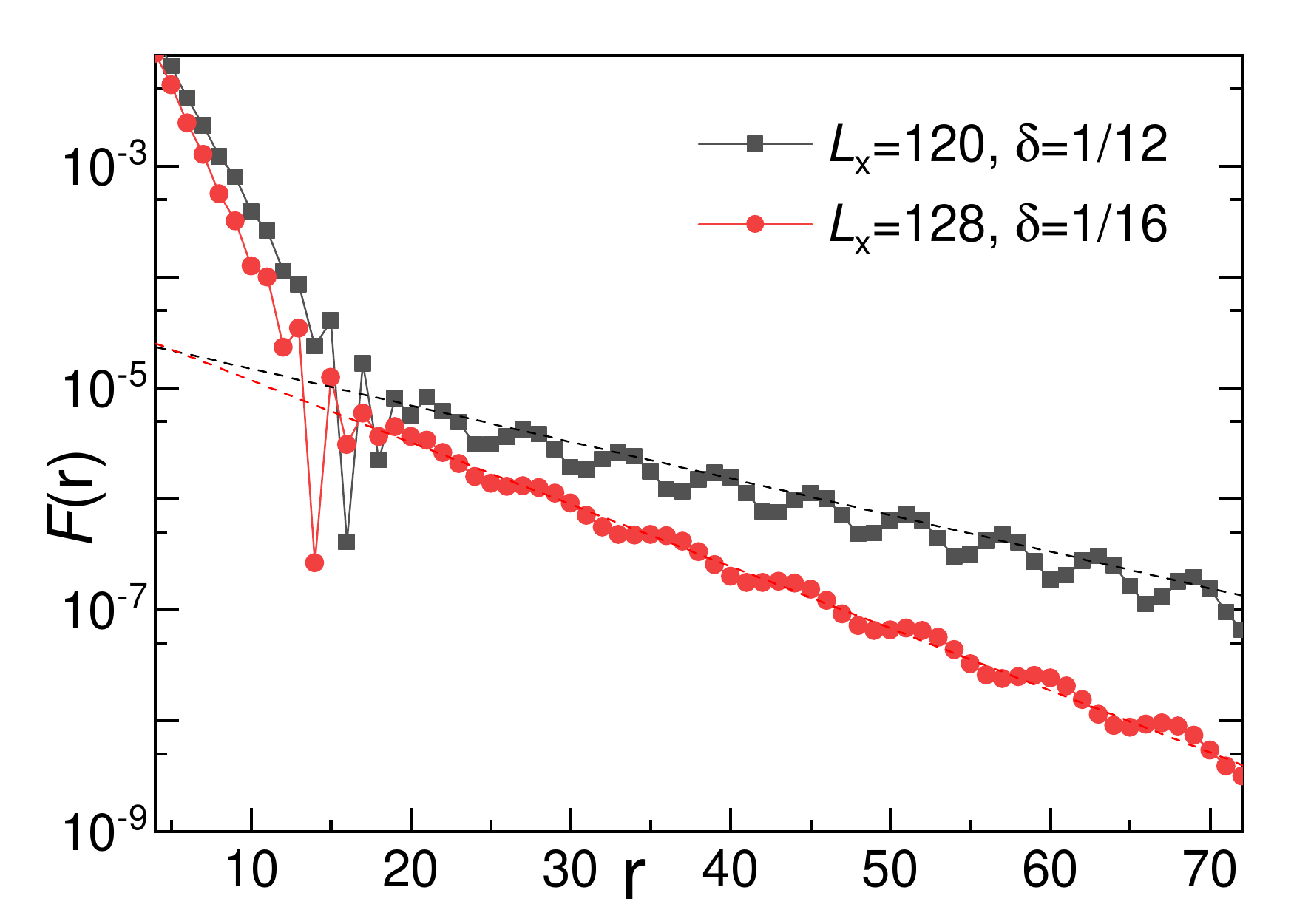}
  \caption{(Color online) Spin-spin correlations $F(r)$ for $L_y=4$ cylinders at $\delta=1/12$ and $\delta=1/16$ on the semi-logarithmic scale. Dashed lines denote exponential fit $F(r)\sim e^{-r/\xi_s}$, where $r$ is the distance between two sites in the $\hat{x}$ direction.}\label{Figs:SpinCor}
\end{figure}

\section{Spin-spin correlations}\label{SM:SpinCor}%
The spin-spin correlations for $L_y=4$ cylinders are provided here. Fig.\ref{Figs:SpinCor} shows the spin-spin correlation $F(r)$ for $L_y=4$ cylinders at $\delta=1/12$ and $\delta=1/16$. The long distance behavior of $F(r)$ is consistent with exponential decay $F(r)\sim e^{-r/\xi_s}$. The extracted correlation lengths are $\xi_s=13.2(1)$ for $N=120\times 4$ cylinder at $\delta=1/12$, and $\xi_s=7.8(1)$ for $N=128\times 4$ cylinder at $\delta=1/18$.

\begin{figure}
\includegraphics[width=\linewidth]{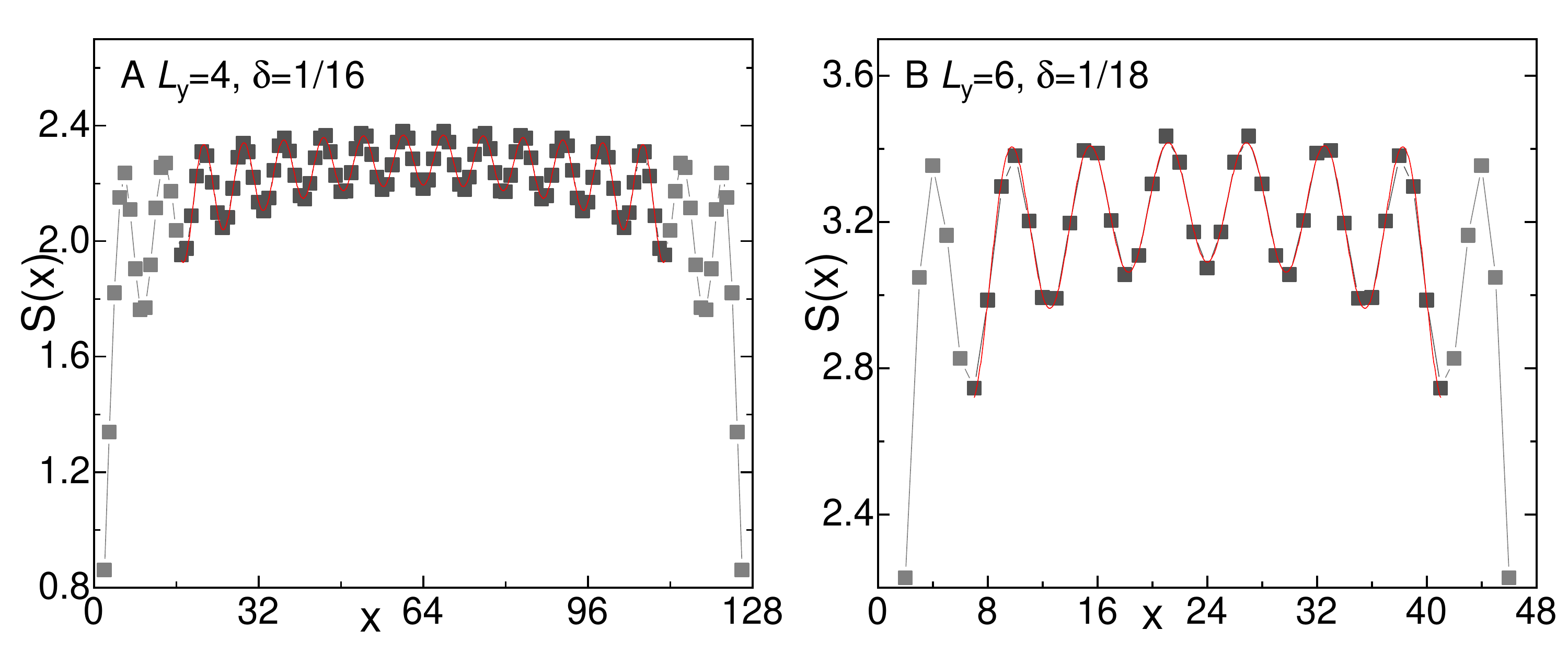}
\caption{Von Neumann entanglement entropy $S(x)$ for (A) $N=128\times 4$ cylinder at $\delta=1/16$, and (B) $N=48\times 6$ cylinder at $\delta=1/18$. Solid lines denote fitting using Eq.\ref{Eq:EE} to extract the central charge $c$. The data points in grey are neglected in the fitting to minimize boundary effect.}\label{FigS:EE}
\end{figure}

\begin{figure}
\includegraphics[width=\linewidth]{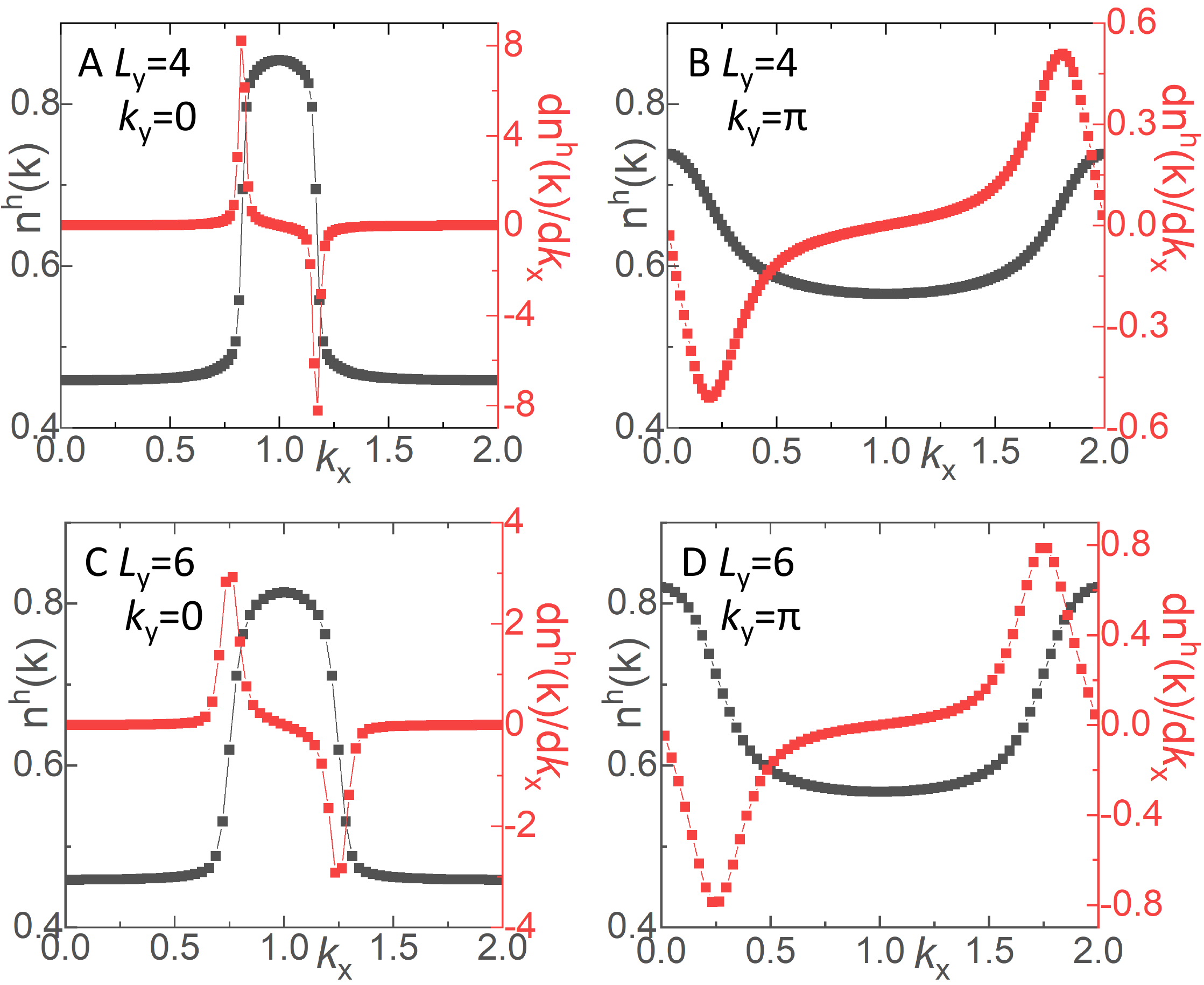}
\caption{(Color online) Hole momentum distribution function $n^h(\mathbf{k})$ at $\delta=1/12$ for $N=120\times 4$ cylinder at (A) $k_y=0$ and (B) $k_y=\pi$, and for $N=48\times 6$ cylinder at (C) $k_y=0$ and (D) $k_y=\pi$, as a function of $k_x$ in units of $\pi$ and the corresponding derivative $dn^h(\mathbf{k})/dk_x$.}\label{FigS:dNk}
\end{figure}

\section{Entanglement entropy and central Charge}\label{SM:EE}%
We have calculated the central charge $c$, which can be obtained by calculating the von Neumann entanglement entropy $S=-\rm Tr \rho ln \rho$, where $\rho$ is the reduced density matrix of a subsystem with length $x$. For a critical 
1D system, it has been established\cite{Calabrese2004,Fagotti2011} that%
\begin{eqnarray}
S(x)&=&\frac{c}{6} \ln \big[\frac{4(L_x+1)}{\pi} \sin \frac{\pi(2x+1)}{2(L_x+1)}|\sin k_F|\big] \nonumber \\
&+&\tilde{A} \frac{\sin[k_F(2x+1)]}{\frac{4(L_x+1)}{\pi} \sin \frac{\pi(2x+1)}{2(L_x+1)}|\sin k_F|}+ \tilde{S},\label{Eq:EE}
\end{eqnarray}
where $\tilde{A}$ and $\tilde{S}$ are model dependent parameters, and $2k_F$ is related to the electron density and, even in a system without a Fermi surface, can be thought of as twice the Fermi momentum.  Examples of $S(x)$ are shown in Fig.\ref{FigS:EE}A and B for $L_y=4$ and $L_y=6$ cylinders, respectively. For a given cylinder of length $L_x$, a few data points in grey close to both ends are excluded in the fitting to minimize the boundary effect. The extracted central charge are $c=1.29(2)$ for $N=120\times 4$ cylinder at $\delta=1/12$, $c=1.03(2)$ for $N=128\times 4$ cylinder at $\delta=1/16$, $c=1.49(5)$ for $N=48\times 6$ cylinder at $\delta=1/12$, and $c=1.37(7)$ for $N=48\times 6$ cylinder at $\delta=1/18$. These values are reasonably consistent with a central charge $c=1$ that is expected for a Luther-Emery liquid by taking into account the finite-size effect.

\section{Hole momentum distribution function}\label{SM:Nk}%
Fig.\ref{FigS:dNk} shows examples of hole momentum distribution function $n^h(\mathbf{k})$ at $\delta=1/12$ for both $N=120\times 4$ and $N=48\times 6$ cylinders. To determine the Fermi momentum for both $k_y=0$ and $k_y=\pi$ bands, the corresponding derivative of $n^h(\mathbf{k})$, i.e., $dn^h(\mathbf{k})/dk_x$, for both $k_y=0$ and $k_y=\pi$ are also shown in the same panels. The peak positions give Fermi momenta $k_0$ and $k_\pi$ in the main text.

\end{document}